\documentclass{fun_article}
\usepackage[english]{babel}
\usepackage{standalone}
\usepackage{import}
\usepackage{xr}
\usepackage{placeins}
\usepackage{afterpage}
\usepackage{lipsum}
\usepackage{amsmath}
\usepackage{siunitx}

\usepackage{bbm}
\author{
	Casper van Elteren\textsuperscript{*, 1, 2, 3}, Vítor V. Vasconcelos\textsuperscript{1,3}, and Mike Lees\textsuperscript{1, 2, 3}\\
\textsuperscript{*}\institution{Corresponding author: caspervanelteren@gmail.com}\\
\textsuperscript{1}\institution{Computational Science Lab, Informatics Institute, University of Amsterdam, The Netherlands}\\
\textsuperscript{2}\institution{Institute for Advanced Study, University of Amsterdam, Amsterdam, The Netherlands}\\
\textsuperscript{3}\institution{POLDER Center, Institute for Advanced Study, University of Amsterdam, The Netherlands}
}

\title{The Paradox of Intervention: Resilience in Adaptive Multi-Role Coordination Networks}
\date{}

\begin{document}

\twocolumn[
 \begin{@twocolumnfalse}
 \maketitle
 \begin{abstract}
 \lettrineabstract{Complex adaptive networks exhibit remarkable resilience, driven by the dynamic interplay of structure (interactions) and function (state). While static-network analyses offer valuable insights, understanding how structure and function co-evolve under external interventions is critical for explaining system-level adaptation. Using a unique dataset of clandestine criminal networks, we combine empirical observations with computational modeling to test the impact of various interventions on network adaptation. Our analysis examines how networks with specialized roles adapt and form emergent structures to optimize cost-benefit trade-offs. We find that emergent sparsely connected networks exhibit greater resilience, revealing a security-efficiency trade-off. Notably, interventions can trigger a "criminal opacity amplification" effect, where criminal activity increases despite reduced network visibility. While node isolation fragments networks, it strengthens remaining active ties. In contrast, deactivating nodes (analogous to social reintegration) can unintentionally boost criminal coordination, increasing activity or connectivity. Failed interventions often lead to temporary functional surges before reverting to baseline. Surprisingly, stimulating connectivity destabilizes networks. Effective interventions require precise calibration to node roles, connection types, and external conditions. These findings challenge conventional assumptions about connectivity and intervention efficacy in complex adaptive systems across diverse domains.
 }
\end{abstract}
\end{@twocolumnfalse}
]
``United we stand, divided we fall'' \cite{DaBoss2013} suggests a principle observed across various scientific domains of the need to coordinate or cooperate. From single-cell organisms \cite{Jones2016, Tero2007, West2016, Brameyer2022} to human societies \cite{Garnett2020, Inoue2019}, collective action walks a tightrope between benefits derived by individuals and the cost required for coordinated interactions \cite{Stewart2014, Su2023, Hegwood2022, Alshamsi2018}.

Networks are used to describe interactions between individuals, with standard interactions including cooperation and coordination, in which individuals often pay a cost to benefit others \cite{Szabo2007}. The conditions for the evolution of cooperation between identical individuals and their connections have been studied using evolutionary game theory \cite{Pacheco2008, Pacheco2006c, Pacheco2006b,Su2023,Allen2024, Szabo2007}. However, individuals are often not identical in their function.

Selective pressures in collaborative networks drive the emergence of specialized roles, as entities optimize their resource allocation through complementary functions rather than redundant capabilities\cite{Ulrich2018, Cooper2018, Merton1934, CazzollaGatti2020}. The resulting complementary roles foster coordination and generate synergistic benefits that evolve to occupy unique niches, each contributing to the collective benefit while optimizing their cost-benefit ratios. These are evident in ecological mutualisms \cite{Raimundo2018, Salles2024}, social sectors interaction \cite{Santos2016, Encarnacao2016}
international trade organizations, and even the structure of clandestine networks \cite{Duijn2014, Kleemans2014, Morselli2007, Morselli2009}. Role differentiation has become a cornerstone feature of these systems, driving their capabilities, efficiency, and resilience.

While the importance of multi-role coordination in complex networks is well established, we lack a comprehensive understanding of how such systems maintain their robustness when adapting to external pressures over time. This gap is particularly evident in systems where the benefits of role specialization must be constantly balanced against increasing costs or risks. How do multi-role adaptive networks evolve to remain resilient under persistent external pressures? This question becomes especially relevant when examining systems that face continuous disruption attempts. Criminal networks offer a compelling example, as they consistently demonstrate remarkable robustness despite coordinated enforcement efforts to dismantle them.

Understanding how to disrupt multi-role networks effectively remains a significant challenge. Interventions can target different aspects of the system, from removing key individuals to disrupting specific relationships or increasing operational costs. Traditional approaches often focus on identifying and removing central actors \cite{Duijn2014, Bright2017}, disrupting specific roles \cite{Morselli2009}, or targeting specific relationships \cite{vanElteren2022}. Yet, the long-term impact of such interventions remains difficult to anticipate, particularly when networks can restructure their relationships and adapt their behaviors.

The limitations of current approaches to control highly adaptive criminal systems are illustrated by Operation Venetic. In 2020, this joint effort by law enforcement agencies from France, the UK, and the Netherlands resulted in more than 800 arrests \cite{Venetic2021}. However, just months later, Dutch authorities reported massive narcotics seizures in Rotterdam \cite{Rotterdam2021}, with rates remaining high \cite{Veiligheid2023}, while UK drug markets quickly rebounded to record levels \cite{UKDrugs2022}. Despite significant intervention, this rapid recovery highlights the urgent need for co-evolutionary models that capture how multi-actor systems evolve in response to external pressures.

We develop and empirically validate a dynamic model of multi-role networks using unique police intelligence data from Dutch criminal organizations. We explicitly capture how entities simultaneously adapt both their behavior and social ties in response to changing costs and benefits. The co-evolution of network structure and individual strategies reveals fundamental mechanisms of network resilience to an array of interventions, applicable across systems from ecological networks to international trade. Then, we calibrate our model with real-world data and demonstrate the bounce-back effects of different interventions.

Our analysis reveals several counterintuitive findings about network evolution and resilience. Contrary to conventional wisdom, increasing network connectivity can destabilize the system. Even though they may increase total potential gains, conditions leading to high link density can shift the equilibrium from a high criminal state to one with minimal criminal activity by amplifying risks relative to potential rewards. Further, we should how interventions that cannot fully disrupt the system lead to temporary increases in performance. These results challenge traditional assumptions about how network disruption affects system stability, suggesting that interventions reducing connections between individuals might actually promote activity growth under low-density conditions.

\afterpage{
\begin{figure*}[t!]
\centering
\includegraphics[width = \textwidth]{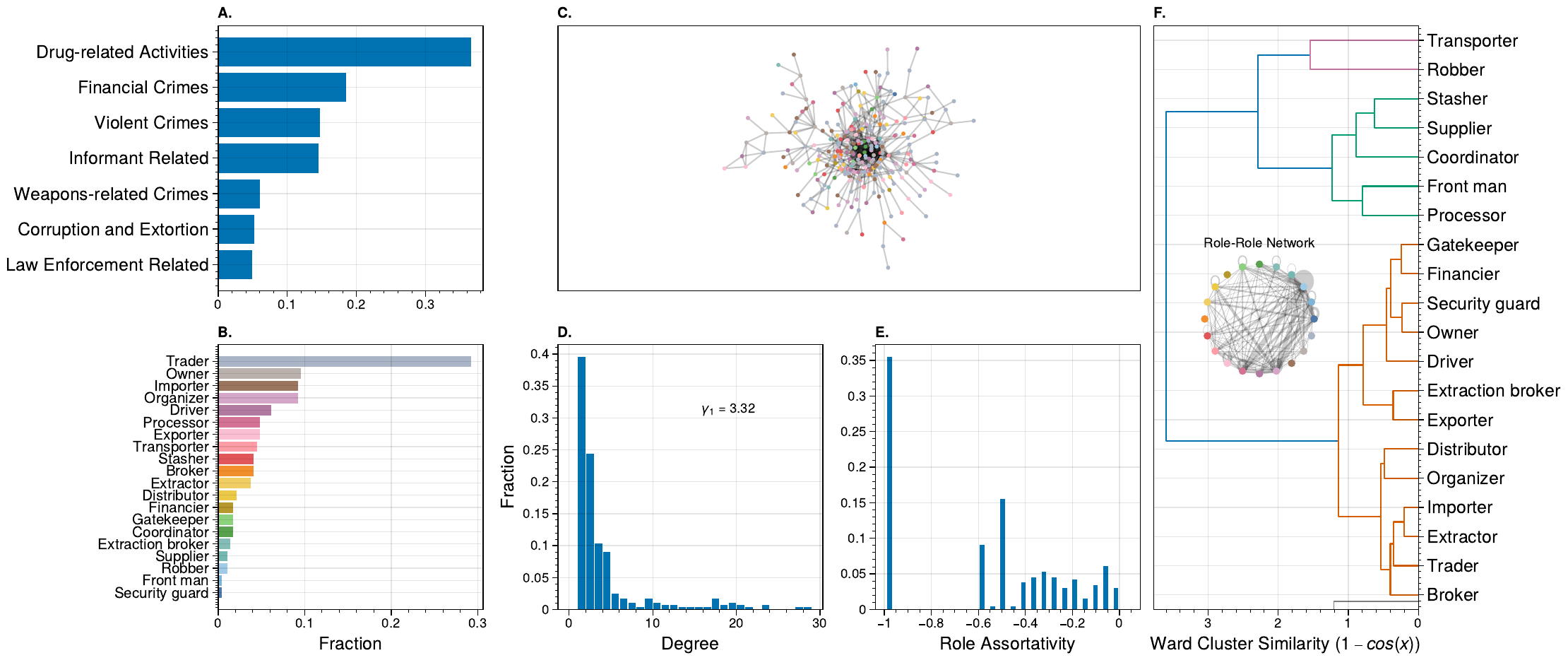}
\caption{\label{fig:fig1} \textbf{The Netherlands is a strategic gateway for European distribution.} The data was provided by the Dutch National Police based on intelligence data in the period 2006-2023. It contains annotated classifications of criminal activities, roles within those, and connections between actors (see Methods for details on data processing and final dataset generation). (\textbf{A}) Relative frequency of criminal activities, highlighting the predominance of illicit drug trafficking. (\textbf{B}) List the fraction of roles in all activities with traders that constitute the majority of actors. (\textbf{C}) The data contains interactions between actors, represented in a network structure; it reveals a large central component surrounded by smaller satellite subgroups. Colors reflect the role assigned to the actor and matching panel B. (\textbf{D}) Distribution of the number of connections per node (degree) in the network, evidencing a right-skewed degree distribution ($\gamma=3.31$). ($\textbf{E}$) The attributed roles display high disassortativity, with agents interacting primarily with others of different roles, suggesting minimal redundancy within the criminal organizations. (\textbf{F}). Hierarchical clustering reveals typical interaction patterns among different roles, showing which roles are more likely to interact with one another.}
\end{figure*}
}

\afterpage{
\begin{figure*}[!t]
\centering
\includegraphics[width=1.\linewidth]{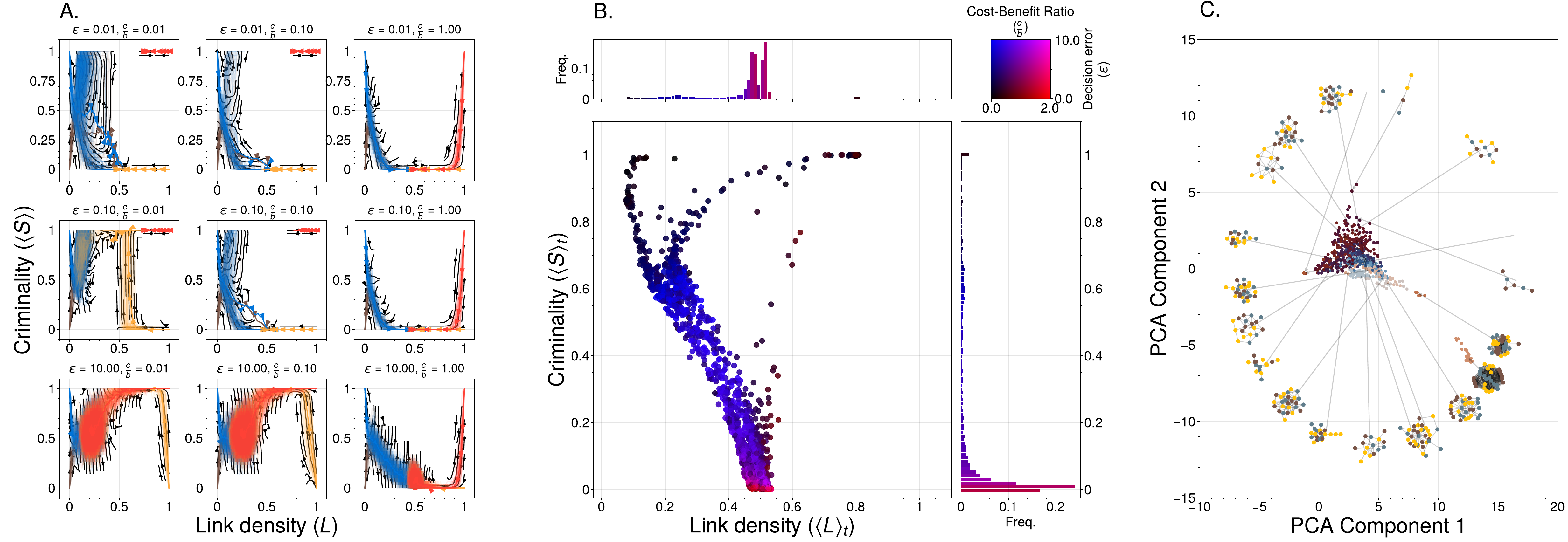}
\caption{\textbf{Emergence and evolution of criminal organizations in a minimal adaptive network model.} \textbf{A}, Network density and criminal activity evolution from four initial conditions (colors), showing multiple attractors that emerge or disappear on cost-benefit ratio and decision noise. Low noise (top) produces distinct criminal/non-criminal attractors; high noise (bottom) yields oscillatory behavior. \textbf{B}, Equilibrium network states ($t=1000$) demonstrate optimization of organizational structure based on cost-benefit ratio and decision uncertainty.  \textbf{C}, K-means clustering analysis of network features reveals diverse organizational structures ($k=20$, see \nameref{app:k-means}). Results from $30$ agents with $3$ equally distributed roles -- shown s colors -- across $10,800$ parameter combinations and one simulation per parameter set.}
\label{fig:fig2}
\end{figure*}
}

\section*{Results}
\label{sec:results}
The Netherlands serves as a crucial gateway for drug trafficking into Europe \cite{Europol2024}. Our analysis of criminal organizations in the Netherlands reveals networks structured around specialized roles, with a primary focus on illicit drug markets and their associated logistics (\cref{fig:fig1}A-C). These organizations exhibit strategic sparsity in their connections, characterized by a right-skewed degree distribution (skewness $\gamma_1 = 3.32$, \cref{fig:fig1}D).

The data shows that criminal actors often possess multiple specialized skills, with individuals holding between one and nine distinct roles across different criminal markets. Using information theory principles, we measure how informative each role is in characterizing an individual's criminal activities. For example, while 'trader' is a common role, its frequent occurrence makes it less informative in distinguishing between individuals. In contrast, a combination of roles like 'trader' and 'electrician' carries more information, as specialized roles like 'electrician' are rarer and, therefore, more salient in characterizing an individual's position in the network. This quantitative approach reveals that individuals strategically form connections with complementary roles while minimizing redundant connections, as evidenced by the negative role assortativity in their ego networks (\cref{fig:fig1}E).

We identified a reduced set of 20 role patterns from the initial set of 99 unique combinations of roles (\cref{fig:fig1}F, see \nameref{app:role_derivation}). This data-driven approach reveals three distinct functional groups: transport (pink), logistics (orange), and management (green), collectively spanning the entire production chain.

\subsection*{Multi-stability and Path Dependency}
Our computational model builds upon the empirically identified roles to explore the dynamics of criminal organization formation. In the model, agents make strategic decisions about both their criminal involvement and network connections based on their roles, and the roles and states of whom they are connected to. This approach allows us to study network configurations and criminal behavior patterns beyond those observed in the empirical data. Agents evaluate the trade-off between potential benefits and risks of criminal activity under uncertainty, parameterized by $\epsilon$. We analyze how the system's behavior varies with both the cost-benefit ratio ($\frac{c}{b}$) and decision uncertainty ($\epsilon$), which jointly shape agents' utility calculations and decision confidence.

We show the coevolution of the fraction of criminals, denoted as criminality, roles, and network link density. In \cref{fig:fig2}(A), we observe the presence of different possible dynamics not only across parameters but within the same parameter set for different initial conditions.

When the cost-benefit ratio ($\frac{c}{b}$) and the decision error ($\epsilon$) are low, multiple dynamic attractors are possible with different levels of stability: i) a high criminality-high density attractor, ii) a low criminality-high density attractor, and iii) an intermediate criminality-low density attractor, each with increasing stability and accessible via different initial conditions. Starting from full connectivity ($L_{t=0}=1$), the networks preserve high interconnection among nodes. In contrast, still starting with all agents adopting criminal strategies but with zero connections ($L_{t=0} = 0$) results in high criminality but low link density. Non-criminal initial states ($\langle S_{t=0} \rangle = 0$) tend to form random networks with a fixed link density, in this case 50\% given the neutral reference payoff of non-criminal actions.

There is a critical threshold for initial link density corresponding to the density of the non-criminal network ($L_{t=0} < 0.5$) that determines whether a criminal organization can form when the cost-to-benefit and decision error approach zero. This threshold plays a crucial role in shaping the emergent network structures.

A single attractor exists for high uncertainty, with higher costs leading to higher link density and lower criminality. The level of success of these organizations depends on the specific cost-benefit and uncertainty.

\cref{fig:fig2}(B) shows, across all parameter settings, how the success or failure of these organizations depends on the considered external factors of cost-benefit and uncertainty but can also be non-unique and sensitive to the system's initial state.

The emerging networks vary beyond their link density. \cref{fig:fig2}(C) illustrates the distinct network structures emerging across cost-to-benefit ratios in a PCA considering an array of network properties, including link density, clustering coefficients, and degree entropy (see \cref{fig:syn_network_properties} for the different properties of the emerging networks). High cost-benefit ratios ($\frac{c}{b} > 1$) result in sparse, fragmented organizations (link density $L < 0.3$, clustering coefficients $< 0.3$). Low ratios ($\frac{c}{b} \ll 1$) lead to denser networks with wider degree distributions (degree entropy $H(k) > 0.7$) and higher average clustering ($C > 0.8$), indicating decentralized, hub-spoke structures. These patterns align with empirical observations of clandestine networks \cite{McMillan2020}, where organizations develop increasingly well-connected structures around key actors when preparing for operations, suggesting similar strategic adaptations in network structure based on operational conditions.

Now that we understand how network structures emerge from different cost-benefit ratios and decision uncertainties, we analyze how these adaptive networks respond to interventions. Using the Dutch Police data, we calibrated the model parameters to reproduce observed network structures (see \nameref{sec:calibration}).

\subsection*{Paradoxical Effects of Interventions on Criminal Networks}\label{sec:interventions}
We systematically evaluated criminal network resilience by simulating three distinct law enforcement strategies: reintegration (increasing agent connectivity through social programs), arrests (isolating agents from the network), and informant recruitment (converting criminal agents to non-criminals while the criminal state of their  connections remains the same).

While we extensively tested these interventions across all parameter combinations (see \nameref{sec:intervention_strategies}), we focus here on the calibrated parameter set that reproduces empirical network structures (\cref{fig:optimization_results}, $\epsilon = 11, \frac{c}{b}=0.11$ with fixed $\lambda = 0.5$). These optimized parameters suggest that criminal organizations in the Netherlands operate with moderate decision uncertainty and face relatively low costs compared to benefits. The high decision error implies considerable mixing between different roles, while the low cost-to-benefit ratio indicates favorable conditions for criminal activities. The model achieves equilibrium networks matching empirical patterns, even with unrestricted rewiring possibilities.

To better reflect real-world constraints, we then restricted agents' rewiring options to local connections (neighbors and neighbors of neighbors), limiting their access to global information. Our analysis reveals intervention outcomes that often contradict intuitive expectations.

\begin{figure*}[!ht]
\centering
\includegraphics[width=.9\linewidth]{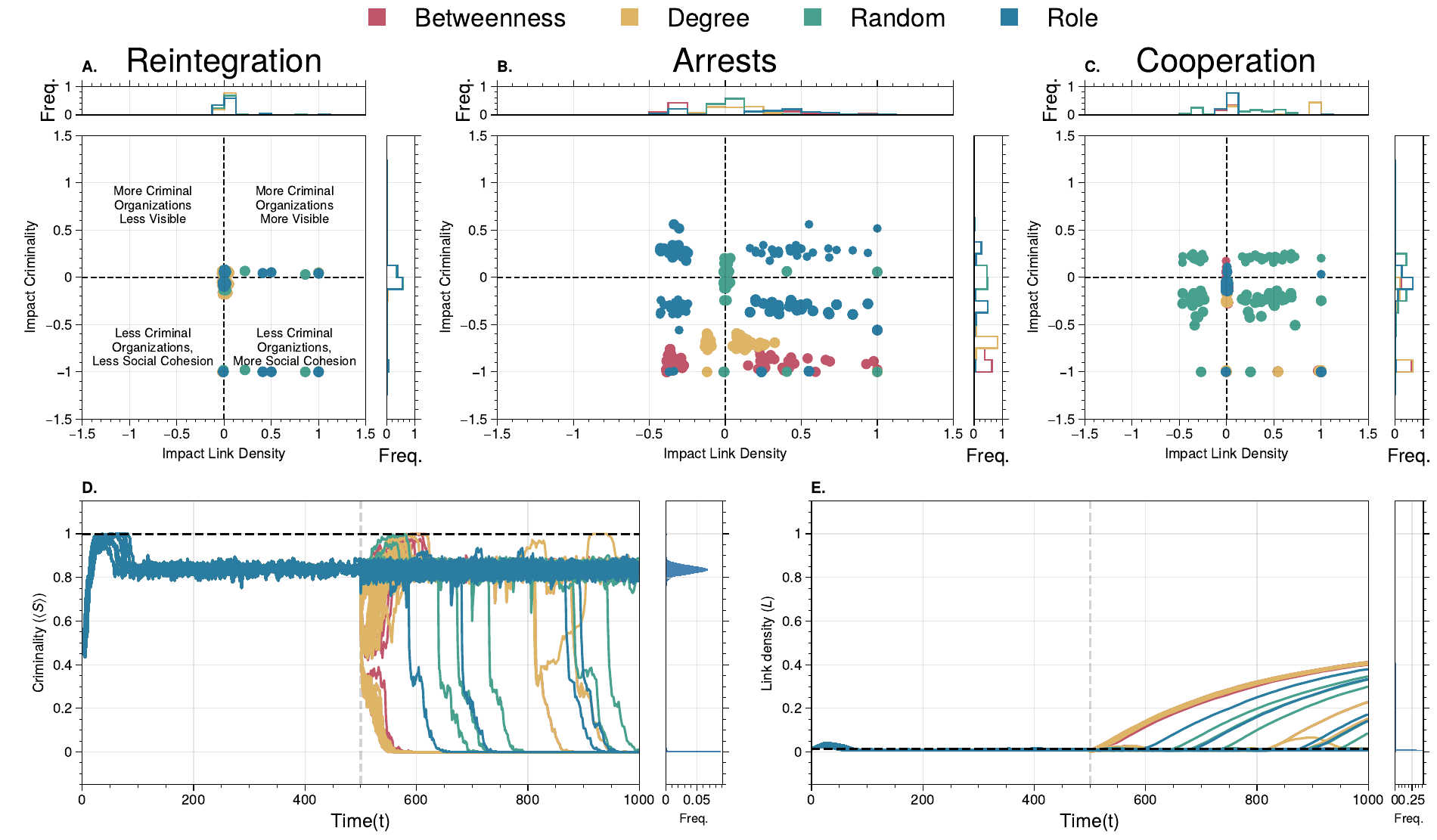}
\caption{\label{fig:fig3} \textbf{Response to interventions. (A) arrests, (B) reintegration, and (C) cooperation.} The effects of interventions (dashed gray line at $t=500$) on criminal networks highly depend on the network's initial state and the intervention strategy. Arrests and informants can lead to increased criminality and decreased network visibility, while reintegration can increase network density. The paradoxical effects of interventions highlight the need for more nuanced, structure-aware crime prevention strategies. Black dashed line indicates the assumed criminality and observed link density of Dutch criminal organizations. The size of the nodes indicates how many interventions ($[1, 2, 10, 20]$ are applied a $t=500$). The bar plots in the panel reflect the data after intervention only.}
\end{figure*}

Criminal networks show varying levels of adaptation post-intervention (\cref{fig:fig3}, top panels). Many networks recover to their pre-intervention state, particularly under reintegration strategies and random arrests. However, role-based arrests and random cooperation produce symmetric deviations from baseline criminal activity, both increasing and decreasing criminality levels. The most effective interventions for reducing criminal activity are arrests targeting agents with high betweenness centrality and degree-based cooperation strategies. While these findings align with previous research on criminal hierarchies \cite{Duijn2014}, our results imply an unexpected pattern: informant recruitment achieves optimal outcomes through random targeting, suggesting criminal networks actively reconfigure roles to maintain resilience.

The co-evolution of these networks shows substantial variability in their post-intervention states, with criminality levels fluctuating up to 50\% from baseline levels (see \cref{fig:fig3}D).

Temporal analysis for the calibrated model revealed complex dynamics. Starting from random configurations that match the density of the empirical network, the simulations demonstrated an initial spike in criminal activity followed by a density adjustment (\cref{fig:fig3} D,E). Although betweenness-based arrest strategies initially reduced both criminality and network density, these effects proved transient, highlighting the network's adaptive capabilities.
When the criminality diminished, the density of the links increased as the non-criminal population approached an Erdos-Renyi configuration ($L = 0.5$). This transition manifested specifically when interventions successfully disrupted the criminal character, resulting in more cohesive but less criminal network structures. The intervention-specific nature of reduced criminality was validated by the absence of such outcomes in pre-intervention simulations.

Our analysis reveals a fundamental tension between network disruption and criminal activity reduction. Strong interventions often triggered compensatory adaptations, leading to increased network resilience. These findings demonstrate that effective and long-term disruption requires sophisticated, systems-based approaches that anticipate and account for the networks' adaptive responses.

\section*{Discussion}\label{sec:discussion}
Our study explores how individual decision-making processes, driven by cost-to-benefit trade-offs and decision certainty, shape the dynamics and collective action in highly adaptive networks, with implications for diverse complex adaptive systems. The observed path dependencies effect demonstrates how initial conditions and external pressures profoundly influence network stability and resilience even in heterogeneous adaptive systems. The observed dynamics between network connectivity and node state highlight a general principle in multi-role coordination systems: functioning that requires balancing collaborative benefits against structural vulnerabilities is robust to various perturbations.

Particularly striking are the paradoxical effects of interventions on criminal networks. We find a ``criminal opacity amplification,'' where targeted disruptions can increase overall criminality while reducing network visibility. This effect challenges conventional law enforcement strategies. It emerges through fragmentation and adaptation, creating more decentralized and fluid criminal landscapes that defy traditional countermeasures. The decrease in link density opens the organization up for potential future attacks; however, it can also make it more difficult for law enforcement to track the organization.

Our co-evolutionary model captures a dynamic security-efficiency trade-off that static network analyses often miss but is often referred to in literature \cite{Morselli2007, Duijn2014,Duijn2016,Bright2017,Kleemans2014, vanElteren2024a}. The ability of criminal networks to reconfigure in response to external pressures mirrors adaptive behaviors observed in legitimate organizational structures and ecological systems. This flexibility principle likely underlies the resilience of many complex collaborative systems, both beneficial and harmful.

While our study focuses on criminal organizations, the insights gained have broader implications for understanding complex adaptive networks across various domains. Our model formulates how networks balance different strategies under external pressures, revealing universal principles of adaptive behavior. These insights could inform the design of resilient business structures in volatile markets or guide the maintenance of biodiversity in ecological networks. The key contribution lies in understanding how adaptive networks respond to interventions—a dynamic that cannot be captured through static network analysis alone.

Despite these promising applications, we acknowledge limitations in our current model. In particular, the model treats all role-to-role interactions as equally weighted and bidirectional, whereas real criminal networks likely exhibit asymmetric relationships between roles. For instance, leadership roles typically exert directional influence over subordinate roles rather than vice versa.

This study represents a rare integration of empirical criminal network data with adaptive agent-based modeling, addressing a theoretical gap in complex adaptive systems but a critical gap in criminological research where data-driven computational approaches remain underutilized. By revealing how local cost-benefit calculations drive emergent network structures, our work provides novel insights into criminal organization resilience that extend beyond traditional statistical analyses. The demonstrated relationship between individual strategic choices and collective adaptation offers valuable insights for policymakers and law enforcement agencies seeking to disrupt criminal networks.

More broadly, our findings contribute to the science of complex adaptive systems by quantifying how targeted interventions can produce counterintuitive outcomes through collective reorganization. This framework for analyzing system-wide responses to local perturbations has particular relevance for understanding organizational resilience, especially in contexts where traditional top-down control is limited or absent. These insights emphasize the critical importance of incorporating adaptive dynamics when designing interventions in complex social systems.

\section*{Methods \& Data} \label{sec:methods}
The model presented here, is an extension of a previous study using fixed graphs. In contrast to that study, we innovate by allowing agents to adjust their connections over time by adding a new rewiring parameter $\lambda$. Furthermore, we used a different sampling technique that is more common in evolutionary game theory. What follows is an in depth description of the model and the data used to generate the results presented in the main text.

\subsection*{Modeling Criminal organizations as a Complex Adaptive System}
At  its core,  we  assume that  a  potential criminal  actor performs  a   cost-benefit  analysis  before   committing  a criminal act. The potential benefits could include financial gain,  enhanced  reputation,  or  the  thrill  of  the  act. In contrast, costs are associated with the risk of being caught or imprisoned. Additionally,  increased criminal  notoriety  may  result  in  potential  harms  from criminal rivalry.

To structure these dynamics, we first outline how agents consider committing a criminal act based on interactions with their nearest neighbors. Then, we describe how agents might decide to change their social
networks by adding or removing connections based on their
current environment.

\subsubsection*{Agent dynamics} \label{sec:agentdynamics}
The model assumes that criminal organizations can be considered to operate similarly to a regular business; agents aim to collaborate with other agents based on the required roles to run a business.  The organization is assumed to be completed by the complete set of required roles harmoniously working together.  Each agent \(i\)  has a state \(s_i \in \{0, 1\}\) that can change based on the nearest-neighbor interactions, whereby a reward \(b\) is obtained when the agents engage in criminal activity. The reward \(b\) is weighed by the cost  \(c\)  that  is  proportional  to  the visibility of an agent in the network.

Formally, the payout for a game consists of the payout of all potential unique games the focal agent $i$ can play with its nearest neighbors. The payout considers the sum of the payoff each agent in the game receives. The payoff for agent $i$ is given by:

\begin{dmath}
\label{eq:payoff_supp}
\pi^{(i)}_s = s \left( b \sum_{O \in \mathcal{O}_i}  \prod_{j}^O s_{j} - c |\mathcal{O}_i|  \right),
\end{dmath}
where $\mathcal{O}_i$ represents all organizations agent $i$ could form with its nearest neighbors and their neighbors, $R$ is the set of all roles required to form a complete organization, $s_{j \in O}^r$ is the state of agent $j$ in organization $O$ having role $r \in R$, $c$ is the cost of being visible in the network.

The potential benefit derived from being part of an organization follows a ``birthday party dynamic" in which the focal agent being updated derives payoffs from organizations formed with its immediate neighbors and the neighbors of neighbors. This results in the focal agent receiving a benefit from a potential partner that it does not directly have a connection with. Consequently, the benefit scales on the order of the average degree, that is the cumulative benefit is proportional $(\langle k \rangle)^2$.

The payoff represents the benefit an agent receives when it connects to a set of actors whose complementary roles enable complete operational capability. The cost is proportional to the number of organizations an agent could be part of but is only incurred when the agent is criminal. Variations of the payoff functions were explored, where the cost was proportional to the (criminal) degree, but the remaining results were similar to the ones presented here and hence will not be further reported on.

The agent considers changing their current strategy $\pi^{(i)}$ with another $\pi{'}^{(i)}$ with probability:

\begin{equation} \label{eq:fermi_supp}
p_{\pi^{(i)} \to \pi{'}^{(i)}} = \frac{1}{1 + e^{-\frac{1}{\epsilon} \left( \pi_{s{'}}^{(i)} - \pi_{s}^{(i)} \right)}},
\end{equation}
where $\epsilon$ is the decision error.

We  further extend  an  agent's capability  by allowing  the agent  to   update  whom  they  are   connected  with.  With probability   \(\lambda\),  each   agent  can   update  that   their connectedness. Each  agent considers  adding or  removing an edge with another random agent: if the agent already possess and edge to  the other agent, it considers  removing it (and adding it vice  versa). The change \(a_i \to  a_i'\) is accepted with probability \cref{eq:fermi_supp}.

Unless  stated  otherwise,  the simulations  were  performed using \(Z=30\) agents with an equal number of agents in each role with a total of \(R=3\) roles. Simulations were performed until the dynamics converged  with an mean squared  error (MSE) \(\le 0.1\)  for   the  duration  of  \(t=100\)   time  steps.  After convergence,  the  system  was  sampled  for  an  additional \(t=100\)  time  steps on  which  the  reported analyses  were performed.  During each simulation step, all agents were updated in random order, yielding a net update of \(t  z  =  30000\) simulation steps.

\subsubsection*{Network dynamics} \label{sec:networkdynamics}
The rewiring  probability,   \(\lambda\),  controls  how   often  agents consider changing their strategy vs. changing their connectivity. For \(\lambda =  0\), agents only switch  from   criminal  to  non-criminal  based   on  their connections   and   strategies,  without   considering   the strategies of  others. At the  other extreme, with \(\lambda  =  \), agents  keep  their current  strategy  and  only attempt  to rewire based on the strategies of others.

In the limit as \(\lambda \to  0\), non-criminal agents yield a payoff of 0 regardless of the strategies of others. Criminals, on the other hand, benefit from connections to other criminal actors, but incur costs proportional to their degree. The full cost $c$ is incurred only when an agent is connected to all others.

This  sets  up  a   trade-off  for  criminals:  forming  new connections    can   increase    benefits   through  novel organizations,    while     incurring    additional    costs $\frac{1}{Z-1}$  per  edge.  Non-criminal agents form a dynamic random network with average degree $\frac{1}{2} Z$. Criminal organizations will attempt to form organizations up to $\frac{M-1}{M} Z$ in the absence of any cost associated with the criminal act. Increasing cost-to-benefit will lead to a reduction in their ability to connect to other agents.

The parameter $\mu$ controls the probability of agents considering connecting to a random other agent. When $mu=1$, agents will  consider removing or adding an edge to a random agent independent of whether it is connected to it already. Conversely, if $\mu=0$ agents will only consider changing their connectivity to their neighbors or neighbors of neighbors. As \(\mu\to0\),  agents tend to  favor connecting to  neighbors of neighbors, resulting  in a more localized  rewiring. In the extreme  case where  \(\mu=0\), agents  only consider  adding or removing  edges to  their immediate  neighbors and  those of their neighbors.

Unless otherwise specified, we set  \(\mu=1\) for the remainder of the experiments. However, see the supplementary information for analysis of other  values of \(\mu\).  The main effects  of varying \(\mu\) are  seen in  scenarios with  low decision  error, where local  information  can  reduce  temporal  fluctuations.  In contrast,   access  to   global   information  reduces   the dependency on  the initial  network structure. As  such, our analysis focuses  primarily on  situations where  actors can interact with  any other actor, assuming  no restrictions on their connections.

\section*{Calibration Procedure using Simulated Annealing} \label{sec:optimization}
We used simulated annealing to optimize the model parameters, ensuring that the model's properties closely match empirical observations while providing insights into system responses to external perturbations. The optimization focused on two key parameters: the decision error ($\epsilon$) and the cost-to-benefit ratio ($\frac{c}{b}$). We fixed the rewiring probability at $\lambda = 0.5$, assuming network restructuring occurs at comparable timescales to strategy updates

The optimization procedure involved 32 independent chains, initialized across the parameter space $c \in [0.001, 2]$ and $\epsilon \in [0.001, 10]$. Each chain was performed for $500$ iterations, with the model evaluated over $1000$ timesteps per iteration and repeated $n = 10$ times for each parameter value to provide confidence in the average cost for that parameter setting -- effectively yielding a maximum likelihood estimate for the model parameters. The quality of the fit was assessed using a cost function that captures both structural and dynamical aspects of the network:

\begin{dmath} \label{eq:cost_function}
F = \sqrt{(\langle S^O \rangle - \langle S^M \rangle)^2 + \frac{1}{Z(Z-1)}\sum_{i, j} (a_{i,j}^O - a_{i,j}^M)^2}
\end{dmath}
where $F$ denotes the distance between the observed quantities (superscript $O$) and model-generated (superscript $M$) quantities: mean criminality ($\langle S \rangle$) and elements of the adjacency matrix ($a_{i,j}$). For the observed data, we assumed a baseline criminality $\langle S^O \rangle = 1$.

Our optimization procedure aimed to demonstrate that our model could successfully replicate the essential structural and dynamical characteristics found in actual criminal networks. While identifying critical features a priori presents a significant challenge, our approach leverages graph edit distance metrics on network links to extract meaningful insights. This method assumes that the observed data contains fundamental truths about the network's structural properties.

By minimizing the distance between observed and simulated networks, we demonstrate that our model can, in principle, generate networks that closely resemble empirical data. However, discrepancies between the model-generated and observed networks may not necessarily indicate model limitations. Instead, these differences could highlight gaps in the observed data due to its inherent incompleteness. Thus, the comparison between modeled and empirical networks serves a dual purpose: it validates our model's accuracy while simultaneously assessing the completeness and quality of the observed data.

The results were stored the runs that minimized the global cost for each chain. The outcomes are represented in \cref{fig:optimization_results}. After the runs, the parameter set was chosen that minimized the overall distance to the real data and used as an input for the interventions. Additionally, synthetic results were generated for a wider set of parameters to assess the model's robustness in response to interventions, please see \nameref{sec:intervention_strategies} for further details.

\section*{Intervention Strategies}
\label{sec:intervention_strategies}
Our study examined the impact of various law enforcement interventions on the dynamics of the criminal network through computational simulations. The intervention process was structured in three phases: warm-up, intervention, and post-intervention.

During the warm-up phase, lasting $1000$ time steps, agents could interact globally, allowing the network to stabilize. At $t = 1000$, we applied interventions with varying intensities, repeating them $n \in \{1, 5, 10, 20\}$ times to simulate scenarios ranging from isolated actions to sustained campaigns. Post-intervention, we restricted interactions to local connections, reflecting increased caution in criminal networks after law enforcement actions. This resulted in setting $\mu=0$ for the remaining $t=1000$ steps.

We targeted agents based on network centrality measures: highest degree, closeness, betweenness, or lowest role assortativity. Three intervention types were implemented:

\begin{enumerate}
    \item \textbf{Reintegration}: The targeted agent received up to n new random connections without changing their criminal state. This method explored how increased local connectivity might influence recidivism.

    \item \textbf{Arrests}: The targeted agent was isolated from the network, simulating incarceration while maintaining their strategy.

    \item \textbf{Cooperation}: The targeted agent's state was changed from criminal to non-criminal. This approach assessed how individual changes impact network stability, with a focus on highly disassortative networks where the removal of a criminal actor could significantly disrupt criminal organizations.
\end{enumerate}

\begin{figure*}
\centering \includegraphics[width=.9\linewidth]{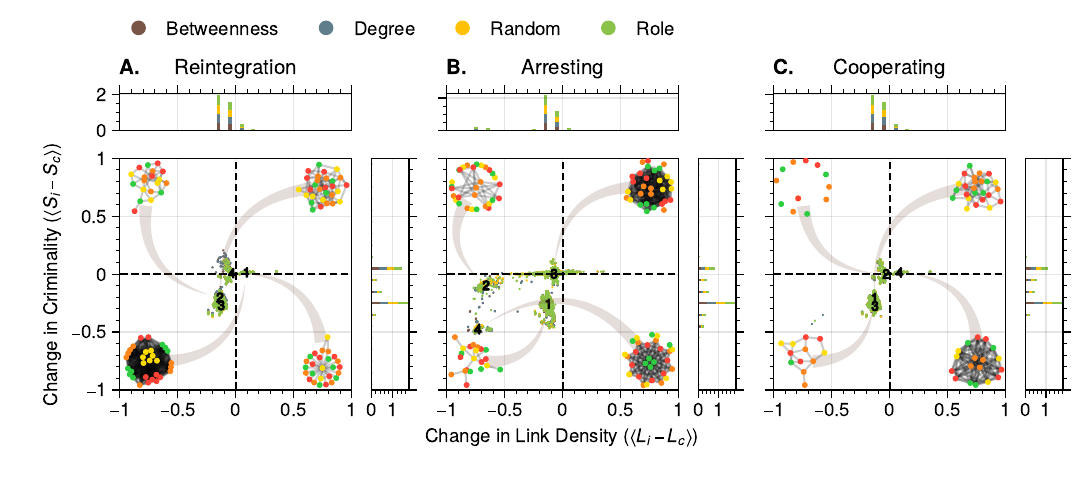}
\caption{\label{fig:synthetic_interventions} \textbf{The effect of emulated law enforcement interventions on synthetic data.} Shown are the results in response to various intervention sizes ($\{1, 2, 5, 20\}$ for $z=30$ for different parameter settings. The four network structures in the corners highlight the network structure closest to the centroid of the cluster indicated by a number. The results show that the effect of the intervention may increase the fraction of criminals post-intervention while decreasing their connectivity -- making them less visible. Furthermore, the results show that the kinds of network structure affected by the interventions may differ. Denser structures are more resilient to interventions. In contrast, decentralized structures are most prone to increasing criminality with reduced link density after intervention, showing additional support for the results found in the main text.}
\end{figure*}

In addition to the intervention performed on the model parameters that best match the data, we performed interventions on synthetic data with the same range as shown in \cref{fig:fig2}. In these simulations, the system consisted of $z=30$ agents and the number of potential roles $R$ was set to 4. The response to the interventions revealed a similar pattern to that of the real data (\cref{fig:synthetic_interventions}, which implies that the model is robust to changes in the number of agents and roles.

By applying these diverse intervention strategies and analyzing their effects on network structure and criminal activity, we aimed to provide insights into the complex dynamics of criminal networks under external pressures. This approach allows for a nuanced understanding of how different law enforcement strategies might reshape criminal landscapes, potentially informing more effective crime prevention policies.

\subsection*{Estimating Intervention Impact} \label{sec:estimating_intervention_impact}
Law enforcement aims to mitigate criminality. Interventions should ideally reduce the number of criminals in the network. However, results from computational interventions may, in fact, cause an increase in criminality rather than reduce it. In some cases, these effects may cause the system to approach a situation in which the system possesses all criminals. The response to interventions can be complex and non-monotonic - we may observe temporary spikes in criminal behavior before the system settles into a new equilibrium. Traditional metrics that focus on average effects or final outcomes may miss these potentially dangerous transient states.
To capture these potentially transient extreme effects, we defined a set of metrics that measure the maximum deviation from pre-intervention behavior. First, we establish a baseline criminality level for each case $i$ by averaging the pre-intervention period from time $c$ to $h-1$, where $h$ marks the start of the intervention:
\begin{align}
\text{baseline}_i &= \frac{1}{h - c} \sum_{t=c}^{h-1} x_{i,t} \label{eq:baseline} \\
\Delta x_{i,t} &= x_{i,t} - \text{baseline}_i \quad \text{for } t \geq h \label{eq:baseline_correction} \\
\Delta x_i^{\text{max}} &= \max_{t \geq h} \Delta x_{i,t} \\
\Delta x_i^{\text{min}} &= \min_{t \geq h} \Delta x_{i,t} \\
\Delta x_i^* &= \underset{k \in \{\text{max}, \text{min}\}}{\text{argmax}} \, |\Delta x_i^k| \\
\Delta x_i^{\text{norm}} &= \frac{\Delta x_i^*}{\max_j |\Delta x_j^*|} \label{eq:imapct}
\end{align}
For each time point after the intervention ($t \geq h$), we calculate the deviation $\Delta x_{i,t}$ from this baseline for the link density and the average criminality. We then identify both the maximum positive deviation $\Delta x_i^{\text{max}}$ and the maximum negative deviation $\Delta x_i^{\text{min}}$. To capture the worst-case scenario, we select whichever of these deviations has the larger absolute magnitude, denoted as $\Delta x_i^*$. Finally, we normalize these effects by the largest absolute deviation observed across all cases $j$, yielding $\Delta x_i^{\text{norm}}$. This normalized metric allows us to compare the relative severity of intervention effects across different scenarios, with values closer to $\pm 1$ indicating more extreme responses to the intervention. For the baseline, we used values from $t>100$ to avoid the transient effects of the initial conditions.

\section*{Data Description} \label{sec:data}
The dataset was provided by the Dutch National Police through an intelligence database system, which serves as a central repository for criminal intelligence information. The dataset contains information on pairwise connections between individuals sourced from criminal investigations: surveillance data, wiretap data, witness- and informant etc, considering individuals only when they occurred across multiple independent sources to ensure reliability. All data is anonymized and no details were provided.

\begin{figure}[htbp]
\centering
\includegraphics[width=.9\linewidth]{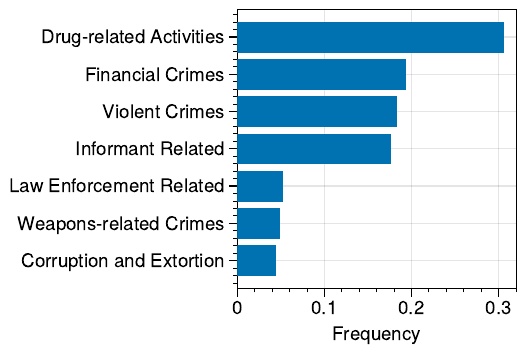}
\caption{\label{fig:criminal_markets} \textbf{In the Netherlands, criminal organizations operate in different markets.}  Illicit drug trade is one of the most dominant criminal markets in the Netherlands. Combined with the financial crime, a picture emerges in which Dutch organizations focus on the distribution and laundering of illicit drugs. The data contains information on the role and activity of individuals in these markets between 2009 and 2023 based on intelligence data obtained from the Dutch National Police.}
\end{figure}

The data spans from 2009 to 2023, with anonymized identifiers and detailed information regarding the roles and activities of individuals in various criminal markets (\cref{fig:fig1}, \cref{fig:criminal_markets}). Data from multiple database sources were combined to provide a cohesive overview of criminal actors in the Netherlands. The database integrates information from multiple law enforcement agencies, including regional police units, the national police, and specialized investigation services.

In total, 20 distinct criminal roles were identified within the network. Each individual could possess up to 9 roles simultaneously, resulting in 99 unique role combinations observed in the data. To ensure data quality, actors with unknown roles were removed from the analysis, reducing the dataset from \num{8995} entries to 603 entries. The final analyzed network consists of 295 individuals connected by 603 edges (see \cref{fig:fig1}), representing verified criminal collaborations. The dataset was provided by the Dutch National Police. The dataset contains information on pairwise connections between individuals sourced based on police intelligence, considering individuals only when they occurred in the joint of different sources.

\section*{Conflict of interest}
\label{sec:org6c0e905}
The authors declare no conflict of interest.

\section*{Data availability}
\label{sec:org415646d}
The data that support the findings of this study are available from the Dutch National Police, but restrictions apply to the availability of these data, which were used under license for the current study, and so are not publicly available. Data are available from the authors upon reasonable request and with permission of the Dutch National Police.

\section*{Code Availability}
The code to  generate the synthetic data  is freely available
at   \url{https://github.com/cvanelteren/complex_criminals}.

\section*{Author contribution}
\label{sec:org1793b31}
\textbf{Casper van  Elteren}:   initial draft, coding,  analysis,  idea
formation.
\textbf{Vítor V. Vasconcelos}: review draft, analysis, idea formation.
\textbf{Mike Lees}: review draft,  analysis, idea formation.

\section*{Acknowledgements}
The authors would like to thank the Dutch National Police for providing the data used in this study.

\section*{Funding}
\label{sec:org37dc282}
This research is supported by  grant Hyperion 2454972 of the
Dutch National Police.

\section*{References}
\label{sec:org171c520}
\printbibliography[heading=none]

\newpage
\appendix
\section*{Appendix} \label{sec:appendix}
\subsection*{Connection to Existing Literature}
\label{sec:orgf25ebe3}
There are  three studies  that looked at  the effect  of law
enforcement  strategies   on  the  disruption   of  criminal
organization. This inherently creates  a tension between two
different  forces;  the  ability  to adapt  to  an  external
stressor.

In  these  studies,  interventions  are  operationalized  by
performing  a  structural  edit   on  the  graph  and  then
simulating  how  the  remaining  in the  agents  replace  or
reconnect  to  remain  operations.  They do  not  model  the
co-evolution of  criminals embedded within a  social network
and implicit economic market.

\subsubsection*{Carley and Tsvetovat}
\label{sec:orge075d63}
The duo  published a series  of articles on  covert networks
\cite{Carley2003,Tsvetovat2014}.  The  topic of  these  papers
involves  building complex  agent  based  models that  could
emulate the behavior or terrorist and criminal organization.
By including a limited  internal state, agents make rational
decisions  on bounded  information.  The  performance of  an
organization to carry out an illicit task is often the focus
of study  while another  explicitly modeled  law enforcement
organization aims to prevent the clandestine organization to
carry out their task.

\subsubsection*{The Relative Inefectiveness of Criminal Disruption}
\label{sec:org48bfe87}
Duijn  and   colleagues  studied   the  effect   of  network
centrality  driven  interventions  and  the  ability  for  a
criminal    network    to    rewire    after    intervention
\cite{Duijn2016}. Different  recovery mechanisms  were studied
in which the agents connected  to a removed individual would
need  to find  a replacement  through their  social network.
Preference  was  given  to,   for  example,  degree,  social
distance, or randomly from a  set of candidates. The results
implied  that  the  average  shortest  path  length  between
individual  reduces  after  intervention. This  is  expected
since  the   rewiring  strategies  are  likely   to  connect
individuals  that are  further  apart and  ensures that  the
network remains connected.

\subsubsection*{Criminal Network Vulnerabilities and Adaptations}
\label{sec:org0f03d18}
Considered the  effect of criminal network  adaptation after
law  enforcement. Interventions  were performed  on a  known
graph  by emulating  law enforcement  intervention based  on
structural  features  of  the networks  (degree  centrality,
betweenness centrality,  and minimum  cut) \cite{Bright2017}.
The effect of intervention were studied with and without the
ability  for  the network  to  adapt.  Adaption occurred  by
considering potential candidates  possessing the right skill
to complete the supply chain for producing a drug.

\subsubsection*{Cost-benefit Modeling and Evolutionary Dynamics}
\label{sec:org6f949b7}
Matjaž  Perc's  research   has  significantly  advanced  the
scientific  understanding   of  criminal   behavior  through
complex       systems       and       network       analysis
\cite{Perc2013a,Perc2015}.   His   studies   revealed   that
criminal  activities  exhibit  clustering  patterns  with  a
small-world network structure, facilitating the rapid spread
of criminal  behavior and making the  network robust against
random  disruptions  but  vulnerable  to  targeted  attacks.
Additionally,  Perc  used  agent-based models  to  show  how
environmental factors, such  as socioeconomic conditions and
law  enforcement  policies,   influence  the  formation  and
sustainability  of  criminal organizations.  These  findings
underscore the importance of  viewing criminal behavior as a
complex adaptive system, advocating  for a holistic approach
to  crime prevention  and  intervention  that considers  the
dynamic  interactions  within  criminal networks  and  their
environments.

\subsection*{Loss of Access to Global Information Reduces Temporal Fluctuations}
\label{sec:org701db63}
In the model, the mixing  parameter \(\mu\), controls the extent
to which  agents can  interact with other  individuals. With
probability \(\mu\),  agents consider connecting with  any other
individual  in the  system  with probability  \(\lambda\). Here,  we
examine the effect on system  dynamics when agents lose this
ability and instead only consider connecting to neighbors of
their  neighbors, thus  constraining  the potential  network
structure.

\begin{figure}[htbp]
\centering
\includegraphics[width=.9\linewidth]{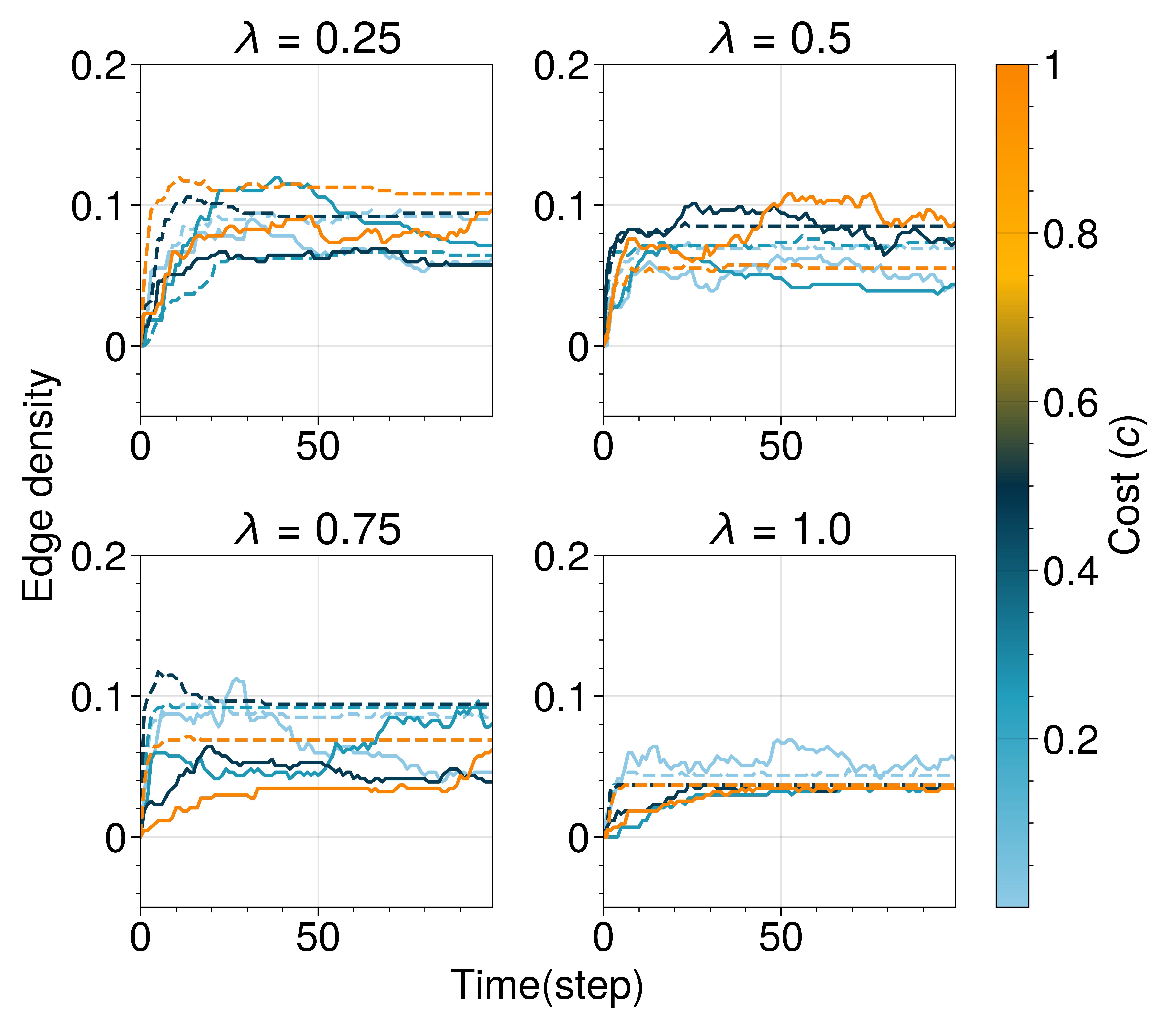}
\caption{\label{fig:loss_of_global_information}The system dynamics show increased fluctuations when agents have access to global information. Shown are the temporal dynamics of the edge density starting form an empty graph with complete criminals having access to global information (solid) and allowing only local information (dashed). The dashed lines stabilize into a regime, lacking temporal fluctuations. Shown are the results for decision error \(\epsilon = 0\).}
\end{figure}

The  system  dynamics  are evaluated  until  equilibrium  is
reached  while  agents  have  access  to  all  other  agents
(\(\mu=1\)).  Then,  agents either  retain  this  access or  are
restricted to  connecting only with neighbors  of neighbors.
The results  show that limiting agents  to local information
reduces    temporal    fluctuations    in    edge    density
(\cref{fig:loss_of_global_information}).

The  findings indicate  that  access  to global  information
introduces some  noise in  the edge dynamics.  This suggests
that while the system can reach  a stable point (as shown in
\cref{fig:fig2}),  this  global  access is  not  necessary  to
maintain  stability. Instead,  it causes  minor fluctuations
over time.

\section*{Deriving Salient Roles from Data}\label{app:role_derivation}
\begin{figure}[htbp]
\centering
\includegraphics[width=.9\linewidth]{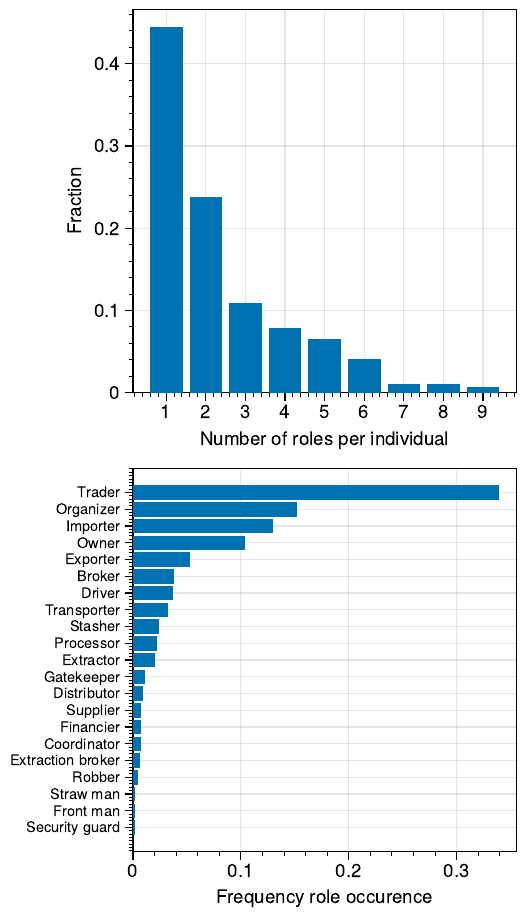}
\caption{\label{fig:network_role} \textbf{Distribution and frequency of criminal network roles in Dutch drug markets.} (top) The number of roles per individual ranges from 1 to 9. The majority of individuals have just one role, with the number of roles increasing as the complexity of required skills increases. (bottom) There is also a notable skew in the frequency of specific roles. The most common role is that of a trader, reflecting the dominant presence of the drug market in the Netherlands.}
\end{figure}

The  value  network was  created  by  separating the  unique
labels into disassortative roles.  We assume that a criminal
organization will aim to  minimize redundancy by interacting
with  roles  different  from  their  own,  resulting  in  an
optimized value  network. The  99 roles are  separated using
techniques from information theory. By leveraging ideas from
natural  language  processing  \cite{Aizawa2003},  we  aim  to
extract the  roles that  convey the most  information, i.e.,
are the most  unique, to determine the full set  of roles an
individual has. In information theory terms, we seek to find
the least number of yes/no questions needed to determine the
full set of roles assigned to an agent starting from a given
role within that subset.

We  define  the  document  \(D\) as  the  collection  of  each
``sentence'' representing the collection  of roles or ``tokens''
that an agent has. The salience of each token is computed as
the shared information between a  token \(t \in T\) and document
\(D\). In other  words, we can compute  the shared information
between the token and the document as the mutual information
\(I(D;T)\)

\begin{dmath}
I(T ; D) = H(D) - H(D | T)  = \sum_{t \in T} p(t|D) p(D) \big(H(D) - H(D | t) \big),
\end{dmath}
where \(H(D)\)  represents the  entropy or uncertainty  in the
document  and \(H(D|T)\)  is the  uncertainty in  the document
given token \(t\). Each agent can then be assigned by its most
salient token, i.e.

\begin{dmath}
a_i = \max_{t \in T} I(t;D).
\end{dmath}
The  resulting  data-driven  value network  is  depicted  in
\cref{fig:fig1}.  Using  the  unique labels,  the  network
display a mostly negative  assortativity for each agents ego
network (\cref{fig:fig1}).

\section*{Model Calibration and Application to Dutch Criminal Networks} \label{sec:calibration}
We calibrated our model's parameters to align with Dutch criminal network data. Using simulated annealing with local gradient checking, we optimized the cost-to-benefit ratio and decision error parameters. The optimization minimized a cost function that quantified discrepancies between the model's network properties and those observed in the data (see methods for details).

The calibrated model successfully reproduces key structural features of the Dutch criminal network. Statistical analysis confirms that the observed data falls well within the model's predictions ($p \ll 0.05$), as evidenced by the right-side area under the curve.

\begin{figure}
\centering \includegraphics[width = 0.45 \textwidth]{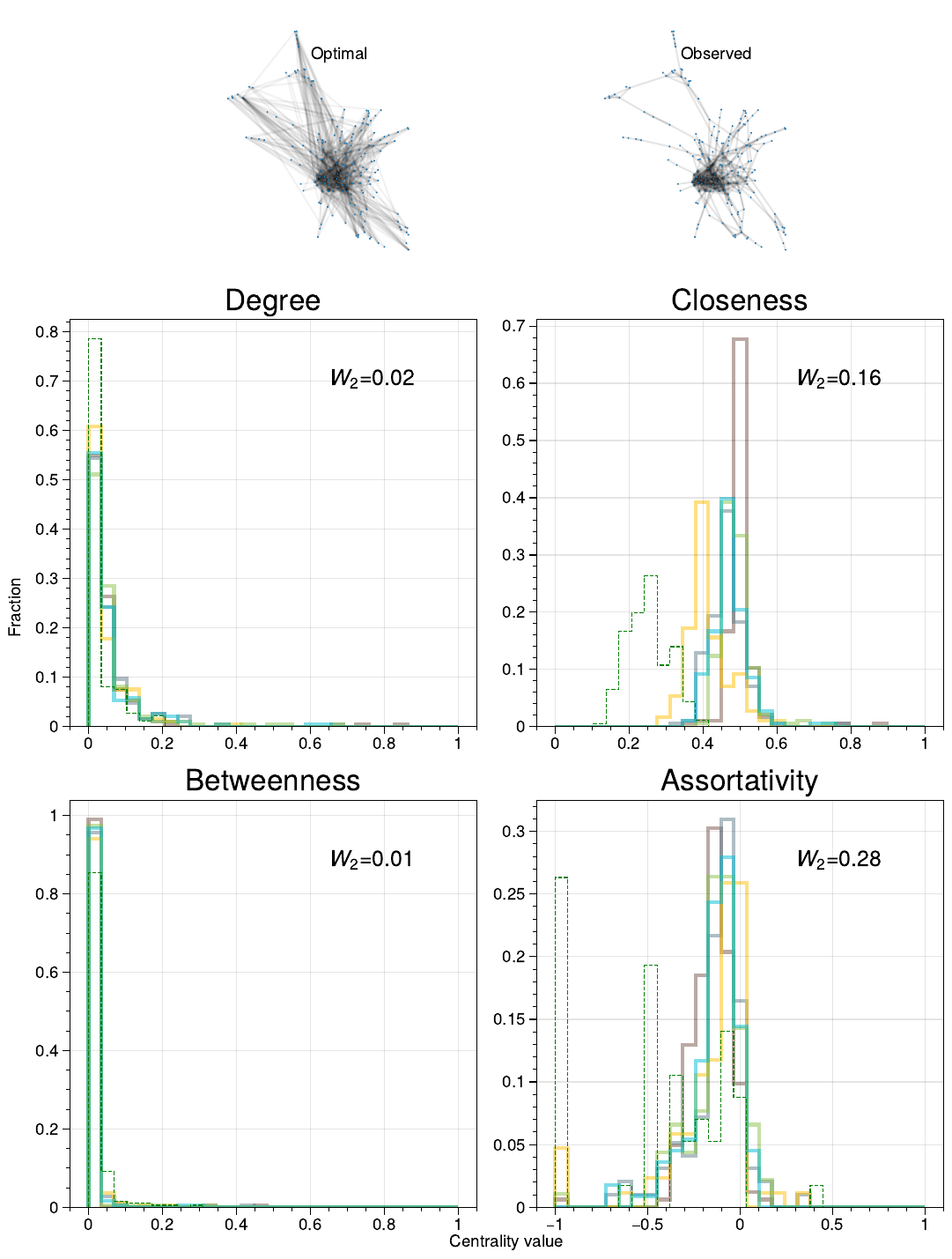} \caption{\label{fig:optimized} \textbf{Comparison of calibrated model to data}. The calibrated model closely resembles the observed Dutch criminal network data (dashed line), as indicated by the low Wasserstein distance ($W_2$, shown for the closest model). Key network features such as degree and betweenness centrality are well approximated, while closeness centrality shows a larger discrepancy. This can be attributed to the higher link density in the optimized model ($L=0.05$), nearly double that of the observed data ($L=0.03$).}
\end{figure}

\section*{Synthetic Data}\label{sec:synthetic_data}
To support the results in the main text, we performed simulations on synthetic data to see how the results generalize. In the main tex we explore how the properties from a real-world network produces a rich set of behavior. To see if this dependency holds for other starting forms, computational experiments were performed on synthetic data with $Z=30$ agents and roles equally occordingly ($Z_i = 10$ for $i \in \{0, 1, 2\}$). Simulations were performed on from four distinct scenarios where the agents were all criminal (or not) or all connected (or not). The results show that the system dynamics are mainly determined by the external conditions of cost-to-benefit ratio and decision error. The rewiring rate was set to $\lambda = 0.5$ unless otherwise stated.

\subsection*{The Zoo of Synthetically Generated Criminal Networks} \label{app:k-means}
\begin{figure}
\includegraphics[width=.9\linewidth]{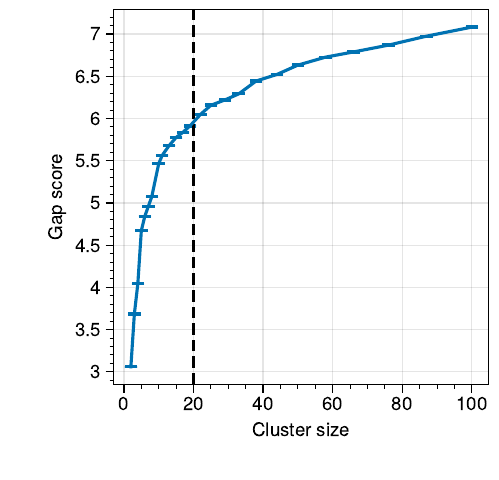}
\caption{\label{fig:gap_score} The network properties of the synthetically generated was separated using the gap score. The optimal number of clusters was found to be $k=20$ with a gap score of $5.90\pm 0.06 2 \sigma$.}

\end{figure}

To show the richness of the proposed generator for criminal collaboration, clustering was performed on synthetic data with $Z=30$ agents. Model comparison revealed that the optimal clustering was achieved for $k=20$ clusters with a gap score of $5.90\pm 0.06 2 \sigma$ (\cref{fig:gap_score}). In total 40 reference distributions were used to create the gap scores. The results are created by simulating the system for $t=1000$ time steps on a grid of $\frac{c}{b} \in \{0.001, \ldots, 2\}$, and $\epsilon =\{0.001, \ldots, 40\}$. The shape of the networks are determined through these exterminal conditions, see main text for further details. Note that our point is not to extract the exact number of clusters, but to show the variety of potential criminal organizations. As such it could be argued that a higher number of clusters could have been chosen as the gap score increases past $20$ clusters.

The bandwidth of potential graphs show a full range from densely connected, decentralized, to a scale-free-like structure of sparse connectivity but some nodes having a high degree.

\begin{figure}
\includegraphics[width=.9\linewidth]{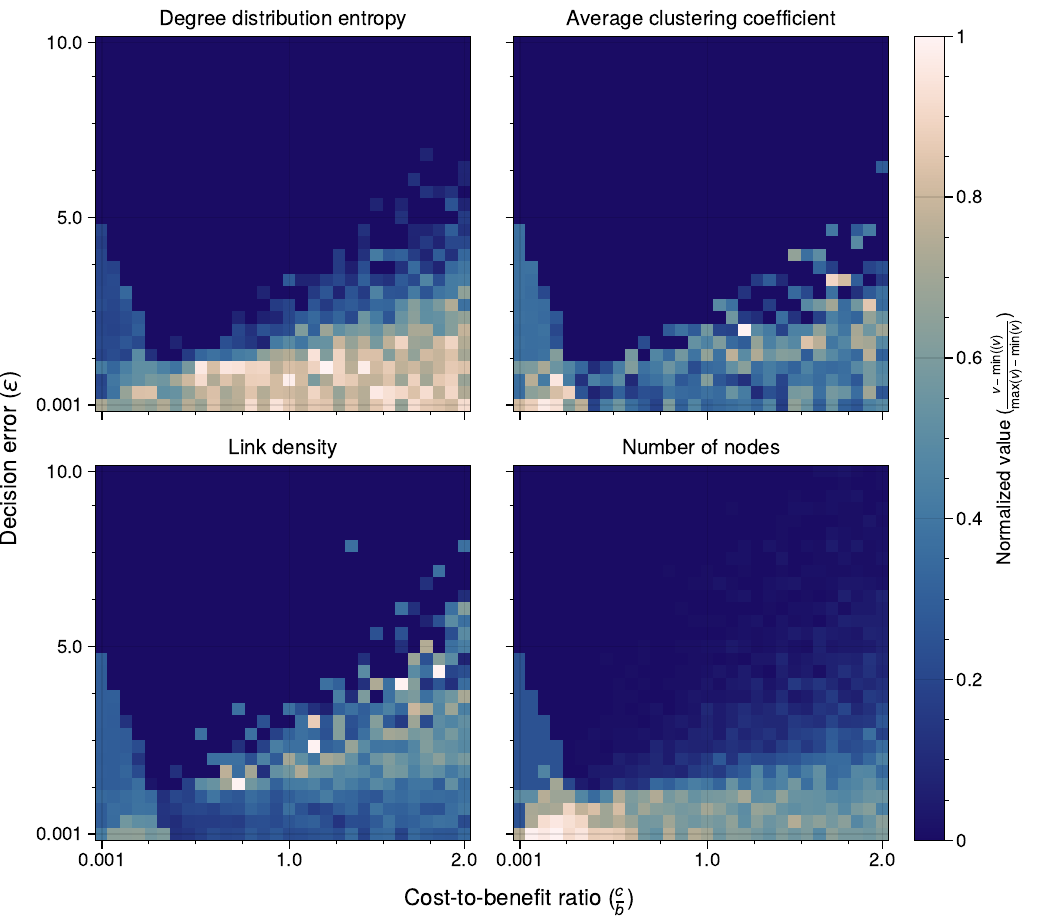}
\caption{\label{fig:syn_network_properties} \textbf{Graphical properties of the synthetic networks.} The entropy of the degree distribution, the average clustering coefficient, link density and number of nodes are shown. Values are normalized within each subplot. The results show that the size of the organization negatively correlates with cost; higher cost reduces the number of members in a criminal organization. The increasing cost, however, increases the variety in the degree distribution.}
\end{figure}

The large varietey of potential criminal organizations structures is reflected in the diverse range of centrality measures. To capture the variety, we show in \cref{fig:syn_network_properties} the normalized values for the entropy of the degree distrbution, the average clustering coefficient, link density and number of nodes. The results show that the size of the organization negatively correlates with cost; higher cost reduces the number of members in a criminal organization. The increasing cost, however, increases the variety in the degree distribution. Together with \cref{fig:fig2}, we can see how the increase in entropy of the degree distribution shifts focus from a decentralized graph to one in which a few nodes have a high degree. This is furhter strengthened by the decrease in clustering coefficient -- lacking short cycles.

\begin{figure*}
\centering \includegraphics[width = \textwidth]{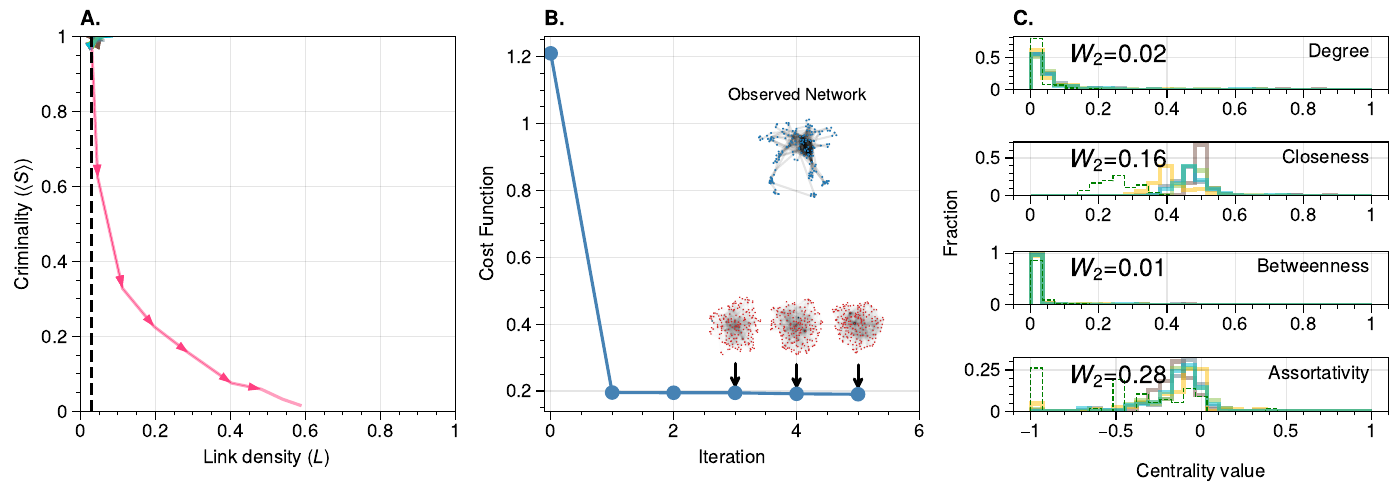}
\caption{\label{fig:optimization_results} \textbf{Optimization results using simulated annealing shows a clear convergence towards the optimal solution.} The results are based on 500 iterations with a cooling rate of $0.995$.}
\end{figure*}

\end{document}